\documentclass{elsart1p}

\newcommand{\pfor}{{\bf for\ }}
\newcommand{\pdo}{{\bf do\ }}
\newcommand{\pto}{{\bf to\ }}
\newcommand{\penddo}{{\bf enddo\ }}
\newcommand{\pwhile}{{\bf while\ }}
\newcommand{\pif}{{\bf if\ }}
\newcommand{\pthen}{{\bf then\ }}

\newcommand{\pendif}{{\bf endif\ }}

\newcommand{\por}{{\bf or\ }}
\newcommand{\pbreak}{{\bf break\ }}

\begin{document}

\begin{frontmatter}

\title{Enhanced molecular dynamics performance with a programmable graphics
processor}

\author{D. C. Rapaport}

\address{Department of Physics, Bar-Ilan University, Ramat-Gan 52900, Israel}

\ead{rapaport@mail.biu.ac.il}

\begin{abstract}

Design considerations for molecular dynamics algorithms capable of taking
advantage of the computational power of a graphics processing unit (GPU) are
described. Accommodating the constraints of scalable streaming-multiprocessor
hardware necessitates a reformulation of the underlying algorithm. Performance
measurements demonstrate the considerable benefit and cost-effectiveness of such
an approach, which produces a factor of 2.5 speed improvement over previous work
for the case of the soft-sphere potential.

\end{abstract}


\end{frontmatter}

\section{Introduction} 

The ability of computers to maintain an exponential performance growth has been
made possible by shrinking component size permitting higher levels of
integration, faster instruction execution and a wealth of hardware capabilities
including cached memory access, multiple instruction units, pipelined
processing, and sophisticated instruction scheduling, to name but a few.
Features leading to higher effective computation speeds that were once confined
to costly high-performance hardware have gradually trickled down to the
affordable CPU chips in current use. Reduced power needs also allow multiple
processor cores to reside on a single chip, a recent notable example being the
graphics processing unit, or GPU (conventional CPUs now also adopt this
strategy). The latest GPUs are fully programmable, and some are even capable of
processing hundreds of separate data streams in parallel. Of the many different
kinds of scientific and engineering computations, those with a more regular data
organization, matrix-vector operations for example, can utilize GPU hardware
very effectively, while the inherent lack of systematically arranged data in,
for example, molecular dynamics -- MD -- simulation, complicates the task of
effective GPU usage.

The availability of optimized computational algorithms is essential for carrying
out MD simulations of large systems over long time intervals. Past efforts
invested in developing hardware-customized algorithms have tended to focus on
high-end supercomputers, with architectures based on vector or parallel
processing, or even both together; the resulting algorithms can be quite
efficient, but introduce additional complexity to overcome hardware constraints.
What is special about the GPU is that it offers high computational capability
while avoiding the cost penalty of other forms of supercomputing since it is a
byproduct of consumer product development (as, indeed, are the microprocessors
powering modern computers in general). Effective GPU utilization also calls for
specialized algorithms, but widespread availability makes it an attractive
platform for MD applications.

The present paper explores the requirements for developing a GPU version of an
efficient, scalable MD simulation for simple fluid systems. Scalability is an
essential characteristic of any algorithm designed for the massive parallelism
intrinsic to present and future GPU designs, and, as will be described in
detail, the approach described here is not subject to the limitations of earlier
efforts that addressed this problem. After a brief outline of the GPU as it
appears from a software perspective, the way the MD algorithms need to be
modified to utilize the hardware features is described, including a short
digression on programming issues specific to the kind of parallelism on which
GPU design is based that, due to their novelty, are still relatively unfamiliar.
Measurements of actual performance and its dependence on various features of the
algorithm are examined, as is the payoff -- actual and potential -- from the
effort invested in the algorithm development.

\section{GPU hardware -- a brief overview}

Graphics have become an integral part of computing, and the demand for increased
capability has resulted in a gradual shift in graphics processor design from
hardwired functionality, via software controlled vertex and pixel shaders, to
the fully programmable GPU \cite{owe08}. The reason a GPU can outperform a CPU,
sometimes by orders of magnitude, is that it is designed to support structured
floating-point intensive computation -- the kind that lies at the heart of the
graphics rendering process -- rather than being optimized to support the high
flexibility demanded from a `conventional' CPU. When provided with a suitable
software interface, the GPU can also be used as a high-performance coprocessor
for non-graphics tasks and, indeed, is a likely building block for the next
generation of supercomputers.

The $\mathrm{CUDA}^{\mathrm{TM}}$ (compute unified device architecture) approach
\cite{nvi09a} is a recent development aimed at simplifying the task of
constructing software to utilize complex GPU hardware without excessive
immersion in the details, while retaining the ability to scale the computation
as more powerful (in particular, increasingly parallel) hardware becomes
available. Conceptually, CUDA operates at a high level of parallelism, and while
in practice concurrency is hardware limited, it exceeds that of a modern
multiple-core CPU by a considerable factor. Parallelism is expressed through
independently executed threads (not to be confused with Unix threads) that are
grouped into blocks; for MD computations, since thread management costs little
in terms of performance, a thread can be assigned to evaluate some quantity
associated with just a single atom, so that there will be as many threads as
there are atoms (without regard for the actual parallelism of the hardware).
This represents the ultimate in fine-grained parallelism.

The ideal program consists of a series of calls by the host CPU to execute
blocks of threads in parallel on the GPU, together with other nonparallel tasks
(hopefully not time consuming) needed to support this effort. Blocks are
processed independently of one another, in parallel to the extent permitted by
the hardware, and then sequentially (strictly speaking, threads are processed in
smaller batches known as warps, a detail mostly invisible to the software).
Although threads do not communicate among themselves, the fact they can access
high-speed shared memory and be mutually synchronized provides a usable hardware
abstraction. If there are data-dependent conditional branches then groups of
parallel threads are executed in series, the groups following alternative paths
with only threads on the path enabled.

Threads all have access to common global memory in the GPU that is separate from
the host memory, and while there is considerable latency involved, a high
bandwidth can result if access is correctly organized in a manner that allows
memory requests by different threads to be coalesced (there are also other
memory spaces, some with faster access, of more limited visibility). For those
classes of problem with well-structured data, e.g., matrix computations, GPU
performance tends to be limited only by the computation rate, while for others,
such as MD simulation, it is the memory access rate that limits performance; the
situation is improved somewhat both by the ability of large numbers of threads
to help conceal memory latency and by the availability of memory caching. In
recognition of the potential usefulness of the GPU as a numerical processor,
hardware improvements are being aimed at eliminating the usability constraints
of earlier designs, examples being the need for increased memory speed and
flexibility, the lack of error-correcting memory, and support for fast
double-precision arithmetic.

\section{MD algorithms}

\subsection{Background}

A typical MD computation entails evaluating forces on atoms (or molecules),
integrating the equations of motion, and measuring various properties
\cite{rap04}. Of these tasks, by far the most intensive is the force evaluation.
In the case of large systems with short-range forces, where each atom interacts
only with a very small fraction of the entire system, the key to efficiency is
the identification of potential interaction partners with a minimum of effort.
This can be accomplished by dividing the simulation region into cells of size
exceeding the interaction cutoff range $r_c$, assigning atoms to cells based on
their current coordinates, and then only examining pairs of atoms in the same or
adjacent cells. This reduces the computational effort for a system with $N_a$
atoms from $O(N_a^2)$ to $O(N_a)$. It is worth noting that systems with
long-range forces can be transformed into an essentially short-range problem;
not doing so is extremely inefficient for large $N_a$, but does provide a good
starting point for learning GPU technique \cite{nyl07} since this naive
$O(N_a^2)$ method is able to use the same efficient, block-organized technique
employed in matrix multiplication \cite{nvi09a}.

A further performance improvement, this time by a multiplicative factor only, is
obtained by using the cell-organized data to construct a list of neighbors that
includes atom pairs with separation $r < r_n = r_c + \delta$, where $\delta$ is
the thickness of a surrounding shell (after enlarging the cells accordingly);
since this list can be guaranteed to include all pairs with $r < r_c$ over
several integration time steps, the work associated with list construction is
amortized over those steps, while the fraction of pairs encountered during the
force evaluations with $r > r_c$ is reduced substantially. The neighbor list is
updated when the cumulative maximum atom displacement reaches $\delta / 2$.

The MD algorithm in this case, after setting up the initial state, involves a
loop containing the following operations \cite{rap04}: (a) the first part of the
leapfrog integration (a half time step update of velocities and a full time step
update of coordinates); (b) if neighbor list updating is required then correct
the coordinates for periodic boundary crossings and do the rebuild; (c) compute
forces and potential energy; (d) the second part of the leapfrog integration (a
half time step update of velocities); (e) evaluate properties such as kinetic
energy and maximum velocity (used to decide when the next neighbor list update
is due); (f) during equilibration adjust the velocities.

This approach is ideal for the conventional CPU; while the neighbor list itself
can be large, depending on $r_n$ and the mean density, only minimal storage is
required to support cell assignment owing to the use of linked lists (see
below). For `unconventional' processors, such as those requiring vector
operations to achieve high performance, or those based on fine-grained
parallelism such as a GPU, the hardware is incompatible with the efficient use
of linked lists, and even simple tabulation of data about neighboring pairs
needs to be rethought.

The problem to be solved is as much one of data organization as it is of
computation; it is an issue that is awkward to accommodate when designing
algorithms for a vector processor, and, to a lesser degree, for a
streamed-multiprocessing GPU. The algorithm designed for the GPU will require
increased storage to avoid the use of linked lists, a change that originated in
attempts to achieve effective vectorization \cite{rap04,rap06}; it will also
entail significantly more computation because Newton's third law will not be
used in order to allow more systematic memory access. Such sacrifices are
justifiable when they contribute to the performance overall. The stages in
converting the computational algorithm to a form that resolves the
incompatibilities are described below. Alternative GPU implementations of MD for
short-range forces are described in \cite{and08,van08}, and the
intermediate-range case in \cite{sto07}; these will be referred to again
subsequently.

Although Ref.~\cite{and08} and the present paper share much in common, there is
an essential difference in the way data is accessed, as described below, that
ensures optimal scaling of the new method as hardware parallelism is increased
in future GPUs, a capability not present in the earlier work. Furthermore, a
relatively large interaction range is needed for the benchmarks reported in
\cite{and08} to achieve efficient hardware utilization, even with the more
limited parallelism offered by past generations of GPUs; reducing the range
below this value, as in the soft-sphere MD example discussed below, leads to a
drastic performance drop, whereas the efficiency of the new approach is not
directly affected by interaction range.

Two interaction potentials are considered. The Lennard-Jones (LJ) potential has
the form
\begin{equation}
u(r_{ij}) = 4 \epsilon \left[ \left(\frac{\sigma}{r_{ij}}\right)^{12} -
\left(\frac{\sigma}{r_{ij}}\right)^{6} \right] \qquad r_{ij} < r_c
\label{eq:lj}
\end{equation}
with a cutoff $r_c$ that must be specified. The soft-sphere (SP) potential is
the same, except that $r_c = 2^{1/6} \sigma$ and a constant $\epsilon$ is added
for continuity. In reduced MD units, length and energy correspond to $\sigma =
1$, $\epsilon = 1$, and atoms have unit mass.

The total number of atoms is $N_a = N_x N_y N_z$, where $N_x$ is the
$x$-component of the size of the ordered atom array in the initial state. The
corresponding edge of the simulation region is of length $L_x = N_x /
\rho^{1/3}$, where $\rho$ is the density. The size of the cell array used for
identifying neighbor pairs is $N_c = G_x G_y G_z$,  with $G_x = \lfloor L_x /
r_n \rfloor$; cells can be indexed both as vectors $\vec{c}$ and scalars $c =
((c_z - 1) G_y + c_y - 1) G_x + c_x$.

\subsection{Simple neighbor lists}

The initial approach, an efficient method for general use \cite{rap04}, provides
a basis for subsequent comparison. Neighbor list construction begins by
assigning atoms to cells, based on coordinates, with cell contents represented
as linked lists of atom indices $q_j$. The first entry in the list for cell $c$
appears in $q_{N_a + c}$, with subsequent entries $q_j$ all having $j \le N_a$;
each $q_j$ is either the identity of the next atom in the list of the owning
cell, or zero if it the last entry. The cell edge is $w_x = L_x / G_x$ and the
simulation region is centered at the origin.

\vspace{6pt}
\begin{tabbing}
\qquad\quad \=\quad \=\quad \=\quad \=\quad \=\quad \=\quad \=\quad \=\quad \=\quad \kill
\> \pfor $c = 1$ \pto $N_c$ \pdo $q_{N_a + c} = 0$ \\[1.8pt]
\> \pfor $i = 1$ \pto $N_a$ \pdo \\[1.8pt]
\> \> $\vec{r}^{\,\prime} = \vec{r}_i + \vec{L} / 2$ \\[1.8pt]
\> \> $c_x = \lfloor r'_x / w_x \rfloor + 1$ (etc.) \\[1.8pt]
\> \> $q_i = q_{N_a + c}$ \\[1.8pt]
\> \> $q_{N_a + c} = i$ \\[1.8pt]
\> \penddo \\[1.8pt]
\end{tabbing}

Enumeration of neighbor pairs employs a set of nested loops, the outermost three
scanning all cells, the next one scanning the offsets between adjacent cells
(only half are needed, 14, including the cell itself), then the two innermost
loops that scan the member atoms of each of the selected pair of cells; atom
pairs in the same cell are only treated once. If the pair separation $\vec{r}$,
after allowing for periodic boundaries, satisfies the distance criterion, $r <
r_n$, the identities of the atom pair are saved in $t'_m$ and $t''_m$, with a
check (not shown) that the size limit of the pair list is not exceeded.
Information about pairs separated by periodic boundaries is packed into $\beta$
(the 27 possibilities are encoded in the 6 high-order bits) and stored along
with the atom identities to avoid the need to repeat the tests each time the
neighbor list is read (this technique is an example of how computation can be
reduced, but it is optional, and only usable if it does not restrict the size of
$N_a$); $\vec{b}_\beta$ is the actual periodic correction to the coordinates
(with components 0, $\pm L_x$, etc.). $N_p$ is the list length.

\vspace{6pt}
\begin{tabbing}
\qquad\quad \=\quad \=\quad \=\quad \=\quad \=\quad \=\quad \=\quad \=\quad \=\quad \kill
\> $m = 0$ \\[1.8pt]
\> \pfor $c'_z = 1$ \pto $G_z$, $c'_y = 1$ \pto $G_y$, $c'_x = 1$ \pto $G_x$ \pdo \\[1.8pt]
\> \> \pfor $k = 1$ \pto 14 \pdo \\[1.8pt]
\> \> \> set $\beta$ and $\vec{b}_\beta$ for periodic boundaries (if any) \\[1.8pt]
\> \> \> $\vec{c}^{\,\prime\prime} = \vec{c}^{\,\prime} +$ cell offset (adjust for periodic boundaries) \\[1.8pt]
\> \> \> $i' = q_{N_a + c'}$ \\[1.8pt]
\> \> \> \pdo \pwhile $i' > 0$ \\[1.8pt]
\> \> \> \> $i'' = q_{N_a + c''}$ \\[1.8pt]
\> \> \> \> \pdo \pwhile $i'' > 0$ \\[1.8pt]
\> \> \> \> \> \pif $c' \neq c''$ \por $i' < i''$ \pthen \\[1.8pt]
\> \> \> \> \> \> $\vec{r} = \vec{r}_{i'} - \vec{r}_{i''} + \vec{b}_\beta$ \\[1.8pt]
\> \> \> \> \> \> \pif $r^2 < r_n^2$ \pthen \\[1.8pt]
\> \> \> \> \> \> \> $m = m + 1$ \\[1.8pt]
\> \> \> \> \> \> \> $t'_m = i'$, \ \ $t''_m = i'' \mid \beta$ \\[1.8pt]
\> \> \> \> \> \> \pendif \\[1.8pt]
\> \> \> \> \> \pendif \\[1.8pt]
\> \> \> \> \> $i'' = q_{i''}$ \\[1.8pt]
\> \> \> \> \penddo \\[1.8pt]
\> \> \> \> $i' = q_{i'}$ \\[1.8pt]
\> \> \> \penddo \\[1.8pt]
\> \> \penddo \\[1.8pt]
\> \penddo \\[1.8pt]
\> $N_p = m$ \\[1.8pt]
\end{tabbing}

The evaluation of the forces $\vec{f}_i$ and total interaction energy $U$
follows; $u(r)$ is the potential energy, Eq.~(\ref{eq:lj}), and $f(r)\, \vec{r}$
the derived force. The computation starts by initializing all $\vec{f}_i = 0$
and $U = 0$, and then treats each of the $N_p$ neighbor pairs. The adjustments
required to account for periodic boundaries are encoded in $\beta$; $B$ is a
mask used to extract the value of $\beta$ ($\bar{B}$ is the complement).

\vspace{6pt}
\begin{tabbing}
\qquad\quad \=\quad \=\quad \=\quad \=\quad \=\quad \=\quad \=\quad \=\quad \=\quad \kill
\> \pfor $m = 1$ \pto $N_p$ \pdo \\[1.8pt]
\> \> $i' = t'_m$, \ \ 
      $i'' = t''_m \,\&\, \bar{B}$, \ \ $\beta = t''_m \,\&\, B$ \\[1.8pt]
\> \> $\vec{r} = \vec{r}_{i'} - \vec{r}_{i''} + \vec{b}_\beta$ \\[1.8pt]
\> \> \pif $r^2 < r_c^2$ \pthen \\[1.8pt]
\> \> \> $\vec{f}_{i'} = \vec{f}_{i'} + f(r)\,\vec{r}$, \ \ 
         $\vec{f}_{i''} = \vec{f}_{i''} - f(r)\,\vec{r}$, \ \ 
         $U = U + u(r)$ \\[1.8pt]
\> \> \pendif \\[1.8pt]
\> \penddo \\[1.8pt]
\end{tabbing}

This simple computation should be contrasted with the corresponding GPU version
developed subsequently. Here, although the $\vec{f}_i$ are read and written
multiple times in no systematic order, the fact that there is just a single
execution thread as well as several levels of data cache for mediating transfers
between CPU and memory should minimize any performance degradation.

\subsection{Alternative neighbor list organization}

The alternative to tabulating neighbor pairs without any specific ordering is to
group the entries according to one member of the pair. Data redundancy is
eliminated, but there is a minor disadvantage that will be indicated below. The
neighbors of atom $i$ are stored sequentially in $t_m$, with $p_i$ pointing to
the first entry; setting the final $p_{N_a + 1}$ ensures the entries for the
last atom are properly terminated. After atoms are assigned to cells, as before,
for each atom $i'$ there are loops over the neighboring cells $c''$ of the cell
$c'$ in which it resides, and then over the atoms $i''$ belonging to $c''$.

\vspace{6pt}
\begin{tabbing}
\qquad\quad \=\quad \=\quad \=\quad \=\quad \=\quad \=\quad \=\quad \=\quad \=\quad \kill
\> $m = 0$ \\[1.8pt]
\> \pfor $i' = 1$ \pto $N_a$ \pdo \\[1.8pt]
\> \> $\vec{r}^{\,\prime} = \vec{r}_{i'} + \vec{L} / 2$ \\[1.8pt]
\> \> $c'_x = \lfloor r'_x / w_x \rfloor + 1$ (etc.) \\[1.8pt]
\> \> $p_{i'} = m + 1$ \\[1.8pt]
\> \> \pfor $k = 1$ \pto 14 \pdo \\[1.8pt]
\> \> \> $\beta$, $\vec{b}_\beta$ and $\vec{c}^{\,\prime\prime}$ (as above) \\[1.8pt]
\> \> \> $i'' = q_{N_a + c''}$ \\[1.8pt]
\> \> \> \pdo \pwhile $i'' > 0$ \\[1.8pt]
\> \> \> \> \pif $i' < i''$ \pthen \\[1.8pt]
\> \> \> \> \> $\vec{r} = \vec{r}_{i'} - \vec{r}_{i''} + \vec{b}_\beta$ \\[1.8pt]
\> \> \> \> \> \pif $r^2 < r_n^2$ \pthen \\[1.8pt]
\> \> \> \> \> \> $m = m + 1$ \\[1.8pt]
\> \> \> \> \> \> $t_m = i'' \mid \beta$ \\[1.8pt]
\> \> \> \> \> \pendif \\[1.8pt]
\> \> \> \> \pendif \\[1.8pt]
\> \> \> \> $i'' = q_{i''}$ \\[1.8pt]
\> \> \> \penddo \\[1.8pt]
\> \> \penddo \\[1.8pt]
\> \penddo \\[1.8pt]
\> $p_{N_a + 1} = m + 1$ \\[1.8pt]
\end{tabbing}

The evaluation of $\vec{f}_i$ is modified to use a double loop over atoms $i$
and over the set of $t_m$ itemizing $i$'s neighbors. The disadvantage is that if
the average number of neighbors is small, the overhead of repeatedly
initializing the inner loop may be noticeable \cite{rap04}.

\vspace{6pt}
\begin{tabbing}
\qquad\quad \=\quad \=\quad \=\quad \=\quad \=\quad \=\quad \=\quad \=\quad \=\quad \kill
\> \pfor $i' = 1$ \pto $N_a$ \pdo \\[1.8pt]
\> \> \pfor $m = p_{i'}$ \pto $p_{i' + 1} - 1$ \pdo \\[1.8pt]
\> \> \> $i'' = t_m \,\&\, \bar{B}$, \ \ $\beta = t_m \,\&\, B$ \\[1.8pt]
\> \> \> compute (as above) \\[1.8pt]
\> \> \penddo \\[1.8pt]
\> \penddo \\[1.8pt]
\end{tabbing}

Each atom pair $(i', i'')$ is considered once during the force evaluation, and
both $\vec{f}_{i'}$ and $\vec{f}_{i''}$ are updated. While the atoms indexed by
$i'$ are accessed sequentially, the $i''$ atoms appear in no particular order.
On the GPU, random read followed by write memory accesses incur a substantial
performance penalty. The alternative is to avoid relying on Newton's third law
by computing $\vec{f}_{i'}$ and $\vec{f}_{i''}$ separately; improved GPU memory
performance more than compensates for the extra computations. The corresponding
modifications to the algorithm are minimal: the loop in the neighbor list
construction (above) over 14 of the neighbor cells is changed to all 27, and the
test for $i' < i''$ is replaced by $i' \neq i''$; only $\vec{f}_{i'}$ is
updated, and the sum over $u$ yields $2 U$. The length of the neighbor list will
of course be doubled.

\subsection{Layer-based neighbor matrix}

A variation of the last approach, based on organizing both the cell contents and
the neighbor pairs as matrices, leads to an algorithm that goes a long way
towards satisfying GPU limitations. Cell assignment is as before, with $c_i$ now
used to record the cell containing atom $i$. Instead of a linked list, cell
contents are organized as a series of layers, as used for vector processing
\cite{rap04}, where layer $l$ includes the $l$th members of all cells (for those
cells with $\ge l$ occupants). Layer assignment is trivial, as shown below;
$l_i$ is the layer containing atom $i$, and $k_c$ serves as a cell occupancy
counter.

\vspace{6pt}
\begin{tabbing}
\qquad\quad \=\quad \=\quad \=\quad \=\quad \=\quad \=\quad \=\quad \=\quad \=\quad \kill
\> \pfor $c = 1$ \pto $N_c$ \pdo $k_c = 0$ \\[1.8pt]
\> \pfor $i = 1$ \pto $N_a$ \pdo \\[1.8pt]
\> \> $k_{c_i} = k_{c_i} + 1$  \\[1.8pt]
\> \> $l_i = k_{c_i}$ \\[1.8pt]
\> \penddo \\[1.8pt]
\end{tabbing}

The standard CPU implementation of this layer assignment algorithm would be just
as shown. When using the layer organization for vector processing, the fact that
the $c_i$ are not unique (cells typically contain multiple atoms) requires the
loop over $i$ to be replaced by a more complicated set of operations consistent
with vectorization. For the GPU version, the increments of $k_{c_i}$ must be
carried out as `atomic' operations to avoid conflict. Newer GPUs support certain
operations of this kind, but they are relatively slow. After testing, it was
decided that this evaluation should be carried out on the host (the remainder of
the work is performed on the GPU); the $c_i$ array (a total of $N_a$ integers)
is copied from the GPU to the host, the values of $l_i$ computed, and the $l_i$
array (of similar size) copied back to the GPU. The maximum of $k_{c_i}$
determines the number of layers in use, $N_l$; this too is most readily
evaluated on the host (on the GPU, a reduction operation -- discussed later --
would be needed).

Once the $c_i$ and $l_i$ are available, the cell-layer occupancy matrix $H_{c,
l}$ can be filled; each atom $i$ contributes a nonzero element $H_{c_i, l_i} =
i$. The row and column indices, $c$ and $l$, specify the cell ($1 \le c \le
N_c$) and layer ($1 \le l \le N_l$); note that while $N_c$ is fixed, $N_l$
varies, and the matrix must be able to accommodate the maximum number of layers
possible.

The final stage is enumerating the neighbor pairs. The results are recorded in
the neighbor matrix $W_{m, i}$, where each column $i$ corresponds to an atom,
and row $m$ specifies the $m$th neighbor of each atom (the order is arbitrary).
This layout, assuming the matrix to be stored in row order, allows the
identities of all $m$th neighbors to occupy successive memory locations
(permitting coalesced access by threads processing individual atoms);
transposing the matrix $W_{m, i}$ reduces performance by $\approx$ 9\%, showing
the sensitivity to memory access issues.

Recording the neighbor pairs involves populating $W_{m, i}$. As before, the
algorithm considers atoms $i'$ sequentially, and for each there are loops, first
over the neighboring cells $c''$ of the cell $c'$ containing $i'$, and then over
layers $l$ to access the neighbor atoms $i'' = H_{c'', l}$. A count of $i$'s
neighbors appears in $m_i$.

\vspace{6pt}
\begin{tabbing}
\qquad\quad \=\quad \=\quad \=\quad \=\quad \=\quad \=\quad \=\quad \=\quad \=\quad \kill
\> \pfor $i' = 1$ \pto $N_a$ \pdo \\[1.8pt]
\> \> $m = 0$ \\[1.8pt]
\> \> $\vec{c}^{\,\prime} =$ cell containing $i'$ (as above) \\[1.8pt]
\> \> \pfor $k = 1$ \pto 27 \pdo \\[1.8pt]
\> \> \> $\beta$, $\vec{b}_\beta$ and $\vec{c}^{\,\prime\prime}$ (as above) \\[1.8pt]
\> \> \> \pfor $l = 1$ \pto $N_l$ \pdo \\[1.8pt]
\> \> \> \> $i'' = H_{c'', l}$ \\[1.8pt]
\> \> \> \> \pif $i'' = 0$ \pthen \pbreak \\[1.8pt]
\> \> \> \> \pif $i' \neq i''$ \pthen \\[1.8pt]
\> \> \> \> \> $\vec{r} = \vec{r}_{i'} - \vec{r}_{i''} + \vec{b}_\beta$ \\[1.8pt]
\> \> \> \> \> \pif $r^2 < r_n^2$ \pthen \\[1.8pt]
\> \> \> \> \> \> $m = m + 1$ \\[1.8pt]
\> \> \> \> \> \> $W_{m, i'} = i'' \mid \beta$ \\[1.8pt]
\> \> \> \> \> \pendif \\[1.8pt]
\> \> \> \> \pendif \\[1.8pt]
\> \> \> \penddo \\[1.8pt]
\> \> \penddo \\[1.8pt]
\> \> $m_{i'} = m$ \\[1.8pt]
\> \penddo \\[1.8pt]
\end{tabbing}

An alternative way of organizing this task, described in \cite{and08} for
neighbor pairs and in \cite{sto07} for forces evaluated directly via the cells,
amounts to having outer loops over cells and their contents, rather than over
the atoms themselves (similar to the original neighbor-list algorithm, above,
but using $H_{c, l}$ instead of $q_i$). Such an approach is designed to utilize
GPU shared memory capability (a subject discussed later), but the disadvantage
is that the number of threads needed per block is determined by maximum cell
occupancy (equal to the layer count $N_l$), a number that can be very small;
this in turn limits the scalability of the computation since it is unable to
benefit from large-scale thread parallelism. The present approach, in which
thread parallelism is limited only by the GPU hardware and not by $r_n$ (which,
in turn, determines cell size and occupancy), is simpler, fully scalable and, as
shown below, exhibits relative performance varying from similar to several times
faster.

Force evaluation is based on $W_{m, i}$. Since evaluation of global sums on the
GPU is nontrivial (see below), the interaction energy of each atom, $u_i$, is
recorded separately, to be combined at a later stage. Note that quantities that
are updated multiple times in the innermost loop, namely $\vec{f}_{i'}$ and
$u_{i'}$, would be held in temporary (register) storage rather than being
written to memory at each iteration.

\vspace{6pt}
\begin{tabbing}
\qquad\quad \=\quad \=\quad \=\quad \=\quad \=\quad \=\quad \=\quad \=\quad \=\quad \kill
\> \pfor $i' = 1$ \pto $N_a$ \pdo \\[1.8pt]
\> \> $\vec{f}_{i'} = 0$, \ \ $u_{i'} = 0$ \\[1.8pt]
\> \> \pfor $m = 1$ \pto $m_{i'}$ \pdo \\[1.8pt]
\> \> \> $i'' = W_{m, i'} \,\&\, \bar{B}$, \ \ 
         $\beta = W_{m, i'} \,\&\, B$ \\[1.8pt]
\> \> \> $\vec{r} = \vec{r}_{i'} - \vec{r}_{i''} + \vec{b}_\beta$ \\[1.8pt]
\> \> \> \pif $r^2 < r^2_c$ \pthen \\[1.8pt]
\> \> \> \> $\vec{f}_{i'} = \vec{f}_{i'} + f(r)\,\vec{r}$, \ \ 
            $u_{i'} = u_{i'} + u(r)$ \\[1.8pt]
\> \> \> \pendif \\[1.8pt]
\> \> \penddo \\[1.8pt]
\> \penddo \\[1.8pt]
\end{tabbing}

\subsection{Additional details}

Most other elements of the computation remain unchanged; only the overall
programming style needs CUDA adaptation, as discussed below. Generation of the
initial state -- consisting of atoms on a cubic lattice of edge size $N_e$ ($N_x
= N_e)$, whose velocities $\vec{v}_i$ have magnitude $\sqrt {3 T}$ ($T$ is the
temperature) and random direction, adjusted so that the system center of mass is
at rest -- is carried out on the host, and the data then transferred to the
GPU. 

For performance reasons associated with memory caching when examining neighbors,
demonstrated later, the sequence in which atoms are stored in memory is
periodically reordered to ensure that the occupants of each cell are kept
together. Reordering involves scanning the most recent version of the occupancy
matrix $H_{c, l}$ in cell order, and it is carried out at regular intervals just
before rebuilding the neighbor matrix (but after periodic boundary adjustments).
A more complicated approach is described in \cite{and08}, where the scan follows
a fractal-like space-filling curve. Only the coordinate and velocity arrays must
be reordered, but not the forces which are due to recalculated. Since reordering
is a comparatively infrequent (and undemanding) operation, the work is carried
out on the host for simplicity. Array elements $k_n$ record the reordered atom
indices and the array $\vec{t}_i$ (of size $N_a$) provides temporary storage
needed while reordering.

\vspace{6pt}
\begin{tabbing}
\qquad\quad \=\quad \=\quad \=\quad \=\quad \=\quad \=\quad \=\quad \=\quad \=\quad \kill
\> $n = 0$ \\[1.8pt]
\> \pfor $c = 1$ \pto $N_c$ \pdo \\[1.8pt]
\> \> \pfor $l = 1$ \pto $N_l$ \pdo \\[1.8pt]
\> \> \> $i = H_{c, l}$ \\[1.8pt]
\> \> \> \pif $i = 0$ \pthen \pbreak \\[1.8pt]
\> \> \> $n = n + 1$ \\[1.8pt]
\> \> \> $k_n = i$ \\[1.8pt]
\> \> \penddo \\[1.8pt]
\> \penddo \\[1.8pt]
\> \pfor $i = 1$ \pto $N_a$ \pdo $\vec{t}_i = \vec{r}_i$ \\[1.8pt]
\> \pfor $i = 1$ \pto $N_a$ \pdo $\vec{r}_i = \vec{t}_{k_i}$ \\[1.8pt]
\end{tabbing}

The last remaining task in the basic MD computation is the evaluation of system
properties requiring sums, or other operations, over all atoms; examples are the
total kinetic and potential energies, based on summing $v_i^2$ and $u_i$, and
the maximum of $v_i^2$ for determining whether the cumulative displacement is
large enough to require updating the neighbor data. Reduction operations of this
kind, trivial on a serial CPU, are more complex on the GPU. The technique --
demonstrated below -- employs a series of partial reductions carried out in
parallel on the GPU, followed by a final reduction on the host.

It is worth reiterating that layer assignment is the only task performed on the
host on a regular basis (its frequency, roughly once every 10-15 time steps, is
measured below); in addition to the minimal amount of computation entailed by
this task, it requires only relatively small amounts of data to be transferred
to and from the host (an array of $N_a$ integers in each direction). Apart from
this, and with the exception of the one-time initialization and the infrequent
atom reordering, the entire computation is executed by the GPU, and there is no
need for any other large, time-consuming data transfers to or from the GPU.

The approach is readily extended to other kinds of MD systems more complicated
than the spherical atoms considered here. For example, only minor modifications
are necessary to allow the GPU to handle multiple atomic species with different
interactions, polymers bound together by internal forces, the velocity-dependent
forces used in modeling granular systems, and even rigid bodies with multiple
interaction sites. There are, however, other extensions of the general MD
approach, such as the incorporation of geometrical constraints or long-range
forces, where efficient GPU implementation requires additional algorithmic
development outside the scope of the present work.

\section{Implementation}

\subsection{Design environment}

A brief discussion of the CUDA features used in this work follows; a more
comprehensive treatment appears in the programming documentation
\cite{nvi09a,nvi09b} and other material available on the Web. Newer and more
advanced hardware than used here includes other features able to improve
performance further; anticipated future gains are likely to come primarily from
increased parallelism, fewer restrictions on memory access, and faster
processing.

The freely available CUDA software environment simplifies development,
especially because of the cooperation between the GPU and host compilers; thus
for simple applications, both the C (or other language) host code and the GPU
functions, written in C, can coexist in the same file for convenience. The
necessary libraries are also provided, so all that is required is a graphics
processor supporting CUDA and its device driver; the present study was carried
out on computers running the Linux (Fedora 11) operating system.

The parallel streaming multiprocessors of the GPU are subject to limitations
similar to those of vector processors regarding data organization and usage. In
the context of MD, since the basic algorithm implicitly allows concurrent
evaluation of the forces between multiple pairs of atoms, careful organization
is required to ensure that individual contributions are combined correctly. This
supplied the motivation for the algorithm redesign described in the previous
section. The problem now is to ensure efficient execution, taking into account
GPU hardware limitations.

The effect of high memory access latency can be reduced by various means, but it
turns out to be the limiting factor, otherwise it would be the Gflop ratio that
determines relative performance. Latency is partially hidden by having a
sufficiently large number of thread blocks, so that while some are awaiting data
from memory others are able to execute. Accessing global memory in the correct
manner allows coalesced data transfers that provide a major increase in
effective bandwidth, e.g., adjacent elements in memory accessed in parallel by a
set of threads. Copying data to shared memory can also improve memory-related
performance, as does the use of texture caching for reading global memory when
coalesced transfers cannot be arranged; however, both these features are of
hardware-limited capacity.

\subsection{Programming}

The CUDA MD implementation entails, for the most part, a few simple C extensions
to support the multithreaded architecture \cite{nvi09a,nvi09b}. The examples
included here are intended to provide a taste of the style and conventions used
in GPU programming. Except where indicated to the contrary, the entire MD
simulation is readily adapted for GPU execution. The first example shows the
function for assigning atoms to cells. The underlying change, here and in other
segments of the computation that run on the GPU, is the removal of the explicit
(outermost) loop over atoms from the function and deriving the atom identity
from the local thread and block indices instead.

The following code, extracted from the host program and simplified, includes
functions that allocate  memory on the GPU and copy an array from host to GPU, a
kernel call with a wait for completion (kernel calls are asynchronous), and an
error check. The function call to \verb|CellAtomAssignGPU| is a request for the
GPU to execute this CUDA function, or kernel, in parallel (the fraction of true
parallelism is hardware dependent). The notation \verb|<<<nBlock, nThread>>>| is
an extension to the C function call mechanism specifying an execution
configuration of \verb|nBlock| blocks, each containing \verb|nThread| threads.
The total thread count, \verb|nBlock * nThread|, ideally equals \verb|nAtom|
($N_a$), although a partially used final block is handled correctly; thus, the
loop over atoms on the CPU is replaced by a single kernel call that processes
them all. The only limit to the total thread count (and also to $N_a$) is GPU
memory; on the other hand, the number of threads per block is limited by
hardware resources and the number of active blocks depends on the memory
requirements per block. Some experimentation may be needed to find the optimal
number of threads per block. Each thread receives the parameters passed in the
call, but otherwise has no access to the host.

The meaning of most variables should be obvious. \verb|r| and \verb|atomInCell|
-- corresponding to $\vec{r}_i$ and $c_i$ -- are arrays in GPU memory. 
\verb|float3| and \verb|float4| specify 3- and 4-component single-precision
values, and \verb|int3| is a 3-component integer. The availability of four
components reflects the graphics origin of the device; their use improves memory
performance, even if the extra padding is space wasted.

\vspace{6pt}
\begin{verbatim}
  cudaMalloc ((void**) &r, nAtom * sizeof (float4));
  cudaMemcpy (r, rH, nAtom * sizeof (float4),
     cudaMemcpyHostToDevice);  
  nThread = 128;
  nBlock = (nAtom + nThread - 1) / nThread;
  CellAtomAssignGPU <<<nBlock, nThread>>>
     (r, atomInCell, nAtom, region, invCellWid, cells);
  cudaThreadSynchronize ();
  if (cudaGetLastError () != cudaSuccess) { ... }
\end{verbatim}
\vspace{6pt}

The kernel \verb|CellAtomAssignGPU| (below) is executed by the threads of the
GPU. Each thread processes a single atom whose unique identity, \verb|id|, is
determined from the following built-in variables: the number of threads per
block, \verb|blockDim.x| (the \verb|.x| suffix arises from the optional
multidimensional indexing of blocks and threads -- not needed here), the
particular block under consideration, \verb|blockIdx.x|, and the thread within
the block, \verb|threadIdx.x| (recall that C uses 0-based indexing). Since
\verb|nAtom| need not be a multiple of \verb|blockDim.x|, a test
\verb|id < nAtom| is included in all kernels. The \verb|__global__| prefix
identifies the function as a CUDA kernel.

\vspace{6pt}
\begin{verbatim}
  __global__ void CellAtomAssignGPU (float4 *r, int *atomInCell,
     int nAtom, float3 region, float3 invCellWid, int3 cells)
  {
    int3 cc;
    int id;

    id = blockIdx.x * blockDim.x + threadIdx.x;
    if (id < nAtom) {
      cc.x = (r[id].x + 0.5 * region.x) * invCellWid.x;
      ...
      atomInCell[id] = (cc.z * cells.y + cc.y) * cells.x + cc.x;
    }
  }
\end{verbatim}
\vspace{6pt}

\subsection{Reduction}

A reduction operation, the simple evaluation of, for example, the sum of a set
of data values, requires careful implementation on a multithreaded GPU to ensure
both correctness and efficiency. Several levels of optimized reduction are
described in \cite{har08}. Acceptable efficiency can be obtained by iterative
reduction on the GPU, halving the number of elements in a block of data until
only one remains (extra effort, not warranted here, can further double the
speed). Since this computation is the most unfamiliar of the changes required
for the CUDA implementation, the second of the software examples demonstrates
the technique. The program fragment shown evaluates the potential energy sum
$\sum u_i$, and is readily generalized to evaluating several quantities at the
same time.

This example also demonstrates the use of shared memory, an important feature of
the GPU for general-purpose computation. Shared memory is not subject to the
high latency of global memory and is available to all threads in a block for the
duration of the block's execution. In the kernel call (below) the third argument
in \verb|<<<...>>>| specifies the amount of shared memory needed (excessive use
of shared memory, a limited resource, can impact performance by reducing
parallelism at the block level). Because it is meaningless for multiple thread
blocks to write to a common memory location in an unsynchronized manner, the
partial result from each block is written to a separate element of the array
\verb|uSumB|, of size \verb|nBlock|, in global memory; the final stage of the
reduction occurs on the host, after copying \verb|uSumB| to the corresponding
host array \verb|uSumH|.

\vspace{6pt}
\begin{verbatim}
  EvalPropsGPU <<<nBlock, nThread, nThread * sizeof (float)>>>
     (u, uSum, nAtom);
  cudaMemcpy (uSumH, uSumB, nBlock * sizeof (float),
     cudaMemcpyDeviceToHost);
  uSum = 0;
  for (m = 0; m < nBlock; m ++) uSum += uSumH[m];
  uSum = 0.5 * uSum;
\end{verbatim}
\vspace{6pt}

When the following kernel is executed there is one thread per atom. The first
task is to copy that atom's $u_i$ into the shared memory array \verb|uSh|.
Subsequent processing reads from and writes to shared memory. Only half the
threads participate in the initial iteration of the \verb|j| loop, and the
number is successively halved down to unity. The synchronization calls to
\verb|__syncthreads ()| are crucial to ensure data is written by all threads
before being read subsequently by other threads. The eventual result is stored
in the element of \verb|uSumB| corresponding to the block.

\vspace{6pt}
\begin{verbatim}
  __global__ void EvalPropsGPU (float *u, float *uSumB, int nAtom)
  {
    extern __shared__ float uSh[];
    int id, j;

    id = blockIdx.x * blockDim.x + threadIdx.x;
    uSh[threadIdx.x] = (id < nAtom) ? u[id] : 0;
    __syncthreads ();
    for (j = blockDim.x / 2; j > 0; j = j / 2) {
      if (j > threadIdx.x) uSh[threadIdx.x] += uSh[threadIdx.x + j];
      __syncthreads ();
    }
    if (threadIdx.x == 0) uSumB[blockIdx.x] = uSh[0];
  }
\end{verbatim}
\vspace{6pt}

\subsection{Texture caching}

The origin of the GPU as a graphics processor is reflected in the ability to use
a cache mechanism for efficiently reading stored textures. This can be put to
general use by binding a data array in GPU global memory to a texture, as in the
following (extended C) host code (the need to invoke graphics capabilities in
this way is rare).

\vspace{6pt}
\begin{verbatim}
  texture <float4, 1, cudaReadModeElementType> texRefR;
  cudaBindTexture (NULL, texRefR, r, nAtom * sizeof (float4));
\end{verbatim}
\vspace{6pt}

Then, rather than the GPU reading coordinates via, for example, \verb|rC =|
\verb|r[id]|, the function call \verb|rC =| \verb|tex1Dfetch (texRefR, id)|
allows the data access to take advantage of the cache. The resulting performance
gain when processing atom pairs (where only the first member of each pair is
accessed sequentially), especially if the data is reordered so that nearby atoms
are also stored in nearby memory locations (as much as possible), will be
demonstrated later. Texture caching is also used for reading the cell-layer
occupancy matrix $H_{c, l}$ when recording neighbor pairs.

\section{Performance measurements}

\subsection{Test environment}

The GPU model used in the tests is the
$\textrm{NVIDIA}^\textrm{\tiny\textregistered}$ Quadro FX770M. This GPU is
designed for laptop computers, an environment subject to strict power and
thermal limitations; thus there are only 32 CUDA stream processors (or four
streaming multiprocessors), 512 Mbytes memory and a 26 Gbyte/s memory bandwidth
(with CUDA level 1.1 support). Other new and recent GPUs have greatly enhanced
performance: several times the number of processors and higher bandwidth,
together with fewer limitations on efficient memory access and support for
double-precision arithmetic.

Performance tests have been carried out for two MD systems. One consists of
atoms interacting with the LJ potential; it is studied for comparison with
earlier work \cite{and08} and has similar parameter settings, $T = 1.2$, $\rho =
0.38$, $r_c = 3$, and $\delta = 0.8$. The other system involves the very
short-range SP potential; it is explored in greater depth, with parameters $T =
1$, $\rho = 0.8$, $r_c = 2^{1/6}$, and (in most cases) $\delta = 0.6$. Other
details, common to both systems, are as follows: The integration time step is
$\Delta t = 0.005$ (MD units), averages are evaluated over blocks of 1000 time
steps, there is an initial equilibration period of 500 steps during which
velocities are rescaled every 20 steps to achieve a mean $T$ close to that
required. Runs are normally of length 6000 steps, which is ample for performance
measurement (except when examining the effect of reordering). Unless stated
otherwise, 128 threads are used, the texture cache is also used, and coordinate
reordering is applied every 100 steps.

\subsection{Size dependence}

GPU performance over a range of system sizes is summarized in Table~\ref{tab1};
times are expressed in $\mu s$ per atom-step (i.e., measured wall clock time per
step divided by $N_a$) and are reproducible with minimal variation (assuming the
GPU is not borrowed by other tasks). The measurements show minimal size
dependence beyond that attributable to variations in the mean number of atoms
per cell $N_a/N_c$. Other performance-related observations include the
following: For the SP case, with the corresponding LJ values shown in
parentheses, there are typically $N_p/N_a =$ 15\,(89) entries in the neighbor
list per atom, $N_l =$ 8--10\,(40--50) layers, and the neighbor list must be
updated every $N_u = $ 11--12\,(14--15) steps. For the largest systems,
fractions $t_n / t =$ 0.38\,(0.35) and $t_f / t =$ 0.37\,(0.58) of the total
computation time, respectively, are used for neighbor list construction and
force calculation (for SP, periodic boundary adjustment, atom reordering, and
cell and layer assignment together account for a fraction 0.03 of the time, even
less for LJ). The computations require GPU storage of $\approx$ 220\,(770)
bytes/atom; the largest systems, with $N_a \approx 8.8 \times 10^5$ ($2.6 \times
10^5$), both need $\approx$ 200 Mbytes. 

\begin{table}
\centering
\caption{Size dependence for soft sphere (SP) and Lennard-Jones (LJ) systems;
the numbers of atoms ($N_a = N_e^3$), cells ($N_c$) and the times per atom-step
($t$) in $\mu s$ are shown.
\label{tab1}}
\vspace{5pt}
\begin{tabular}{rrrrc}
\hline
    & $N_e$ &  $N_a$  &  $N_c$  &  $t$    \\
\hline\\[-10pt]
 SP &  32  &   32768  &   8000  &  0.072  \\
    &  40  &   64000  &  13824  &  0.070  \\
    &  48  &  110592  &  27000  &  0.067  \\
    &  56  &  175616  &  39304  &  0.068  \\
    &  64  &  262144  &  64000  &  0.068  \\
    &  72  &  373248  &  85184  &  0.067  \\
    &  80  &  512000  & 125000  &  0.067  \\
    &  88  &  681472  & 157464  &  0.069  \\
    &  96  &  884736  & 216000  &  0.067  \\
\hline\\[-10pt]
 LJ &  32  &   32768  &   1000  &  0.266  \\
    &  40  &   64000  &   2744  &  0.240  \\
    &  48  &  110592  &   4096  &  0.270  \\
    &  56  &  175616  &   8000  &  0.243  \\
    &  64  &  262144  &  10648  &  0.254  \\
\hline
\end{tabular}
\end{table}

\subsection{GPU vs CPU and other performance comparisons}

The most important result is the magnitude of the performance improvement
relative to a more conventional CPU. The C version of the layer-matrix program
(corresponding to the CUDA version) was run on a Dell Precision 470 workstation
with a 3.6GHz Intel Xeon processor, similar to, but slightly faster (nominally
1.2x) than that used in \cite{and08}, and compiled with maximum optimization.
Timings appear in Table~\ref{tab2}, together with the speedup factor. The gain
is impressive, and in the case of LJ, consistent with \cite{and08} which used a
GPU (NVIDIA GeForce 8800 GTX) with 4x the number of processors (128 vs 32) and
over 3x memory bandwidth (86 vs 26 Gbyte/s, reflecting a correspondingly wider
memory interface); the approach in \cite{and08} is itself several times faster
than \cite{van08} (which used cells but not neighbors) on the same model GPU.

\begin{table}
\centering
\caption{Performance comparisons -- GPU vs CPU.
\label{tab2}}
\vspace{5pt}
\begin{tabular}{rrrrc}
\hline
      & $N_e$ & \multicolumn{2}{c}{$t$} & $t_{CPU} / t_{GPU}$ \\
      &       & {\small CPU} & {\small GPU} &  \\
\hline\\[-10pt]
 SP   &   64  &  0.795  &  0.068  &  11.7   \\
 LJ   &   40  &  4.480  &  0.240  &  18.7   \\
\hline
\end{tabular}
\end{table}

Another important comparison addresses the performance of the old
(Ref.~\cite{and08}) and new methods as $r_c$ is varied, particularly since the
earlier work considered an LJ system with only a single, relatively large value
of $r_c$. The algorithm used to tabulate neighbors in \cite{and08} is readily
incorporated into the present program (after correcting a few minor errors)
since similar matrix organization is used to represent cell and neighbor data.
The timing measurements appear in Table~\ref{tab3}. For the largest $r_c$ the
two methods exhibit similar performance, as indicated above, but as $r_c$ is
lowered (using values from \cite{rah64,ver67}) the performance figures begin to
favor the new approach, culminating in an almost 2.5x gain in the SP case. The
improvement is due solely to the modified approach to neighbor tabulation
(which, unlike the old method, does not require a large $r_c$ to benefit from
thread parallelism). For the SP system, the time fractions used for neighbor
list construction are $t_n/t =$ 0.34 and 0.74 for the new and old methods,
respectively; the latter value reveals how this portion of the computation
dominates in the old method. The present new approach is expected to show
further improvements in performance when used with more advanced (current and
future) GPUs incorporating greatly increased parallel capability.

\begin{table}
\centering
\caption{Performance comparisons -- new vs old (\cite{and08}) algorithms.
\label{tab3}}
\vspace{5pt}
\begin{tabular}{crrrc}
\hline
             	&$N_e$& \multicolumn{2}{c}{$t$} & $t_{old} / t_{\vphantom{d}new}$ \\
             	&     &  {new}	&  {old}  &	   \\
\hline\\[-10pt]
 LJ $r_c=3.0$   & 48  &  0.270  &  0.254  &  0.94  \\
 LJ $r_c=2.5$   & 48  &  0.176  &  0.185  &  1.05  \\
 LJ $r_c=2.2$   & 48  &  0.137  &  0.164  &  1.20  \\
 SP	        & 64  &  0.068  &  0.168  &  2.47  \\
\hline
\end{tabular}
\end{table}

There is a performance loss associated with the alterations that were made to
the MD algorithm. A series of measurements for each of the versions of the
neighbor list and force computations conducted on the $N_e = 48$ SP system,
subsequently referred to as {\em S}, were run on an Intel T9600 CPU (2x the
speed of the 3.6GHz Xeon) belonging to the laptop with the GPU (HP EliteBook
8530w). Relative to the base algorithm that used a list of neighbor pairs,
grouping the neighbors by atom (with the extra inner loop) increases the time by
1.16x, relinquishing Newton's third law a further 1.42x, and the use of the
layer matrix another 1.17x; the cumulative increase is 1.92x. Thus, in order to
reap the benefits of the GPU hardware it is necessary to work with an algorithm
roughly half as efficient as the original, but the loss is more than justified
by the net performance gain. A similar situation arose when adapting MD for
vector processing; whether some of the other changes made to aid vector
efficiency might be helpful here remains to be investigated.

The more processing the GPU can apply to data retrieved from memory (the
`arithmetic intensity') the greater the expectation for improved performance.
With short-range MD, and the SP problem in particular, atoms have few
interaction partners, a fact that offers a difficult challenge for the GPU when
compared to a CPU. Given the positive outcome of the comparison, despite the
unfavorable nature of the problem, the future of the GPU-based approach appears
promising.

\subsection{Energy conservation}

Energy conservation is an essential requirement for any MD simulation in the
microcanonical ensemble. It is simple to test whether the fact that the GPU used
here is limited to single-precision arithmetic affects this capability.
Table~\ref{tab4} compares the values at the beginning and end of a longer run
(excluding the first 1000 steps that are influenced by velocity rescaling during
equilibration). The measured drift for system {\em S} is 1 part in 5000 over
$10^5$ steps, a more than acceptable value.

\begin{table}
\centering
\caption{Test of energy conservation showing mean total and kinetic energy per
atom, $\langle e_{tot} \rangle$ and $\langle e_{kin} \rangle$, averaged over the
1000 steps preceding $N_{step}$.
\label{tab4}}
\vspace{5pt}
\begin{tabular}{rcc}
\hline
$N_{step}$  & $\langle e_{tot} \rangle$  &  $\langle e_{kin} \rangle$   \\
\hline\\[-10pt]
   2000  & 2.3260603  &  1.5000234   \\
 100000  & 2.3265371  &  1.5005983   \\
\hline
\end{tabular}
\end{table}

\subsection{Parameter dependence}

Table~\ref{tab5} shows the dependence on $\delta$, the shell thickness, for
system {\em S}. As $\delta$ increases fewer cells are used, and there are
increases in both the layer count $N_l$ and the interval between neighbor list
updates $N_u$. Computation time reflects the decreasing amount of work required
for the neighbor lists and the corresponding increase for the forces (although
the overall variation is fairly small over the $\delta$ range considered), with
a minimum in the vicinity of $\delta = 0.6$ as used previously.

\begin{table}
\centering
\caption{Dependence on shell thickness ($\delta$) for system $S$; the values
shown are the number of cells ($N_c$), the mean number of neighbors ($N_p /
N_a$), the maximum number of layers ($N_l$), the mean number of steps between
neighbor updates ($N_u $), the fractions of computation time devoted to neighbor
processing ($t_n/t$) and force calculation ($t_f/t$), and the time per atom-step
($t$).
\label{tab5}}
\vspace{5pt}
\begin{tabular}{cccrrccc}
\hline
$\delta$ & $N_c$ & $N_p/N_a$ & $N_l$ & $N_u$   & $t_n/t$ & $t_f/t$ & $t$    \\
\hline\\[-10pt]
  0.4    &  32768  & 11.57   &   7   &   8.1   & 0.425   &  0.295  &  0.070 \\
  0.5    &  27000  & 13.33   &   8   &  10.1   & 0.378   &  0.337  &  0.069 \\
  0.6    &  27000  & 15.44   &   8   &  12.0   & 0.331   &  0.386  &  0.067 \\
  0.7    &  21952  & 18.12   &   9   &  14.0   & 0.310   &  0.425  &  0.070 \\
  0.8    &  17576  & 21.64   &  11   &  15.9   & 0.292   &  0.462  &  0.075 \\
\hline
\end{tabular}
\end{table}

The number of threads per block, $N_T$, is a runtime parameter that must be
determined empirically, and will vary with GPU capability as well as with
problem type and size; $N_T$ should exceed and be a multiple of the number of
GPU processors (here 32). Table~\ref{tab6} shows the performance of system {\em
S} with different $N_T$; processors must be kept busy, while not overutilizing
resources available to the threads. The choice of 128 threads used throughout
the study appears justified.

\begin{table}
\centering
\caption{Dependence of time per atom-step on thread count ($N_T$).
\label{tab6}}
\vspace{5pt}
\begin{tabular}{rc}
\hline
 $N_T$   & $t$    \\
\hline\\[-10pt]
   32	 & 0.097  \\
   64	 & 0.067  \\
  128	 & 0.068  \\
  256	 & 0.085  \\
\hline
\end{tabular}
\end{table}

The importance of reordering based on the atom coordinates as a means to
improving memory access times was mentioned earlier. Table~\ref{tab7} shows the
effect of varying the nominal number of steps $N_R$ between reorderings (the
operation is carried out when the next neighbor list rebuild falls due) for
system {\em S}. Frequent reordering clearly makes an important contribution to
performance; the value $N_R = 100$ is a reasonable choice.

\begin{table}
\centering
\caption{Dependence of time per atom-step on reorder interval ($N_R$).
\label{tab7}}
\vspace{5pt}
\begin{tabular}{rc}
\hline
 $N_R$    &  $t$    \\
\hline\\[-10pt]
  50      &  0.066  \\
  100     &  0.067  \\
  250     &  0.070  \\
  500     &  0.075  \\
  1000    &  0.081  \\
  2000    &  0.094  \\
\hline
\end{tabular}
\end{table}

Table~\ref{tab8} shows the consequences of failing to reorder, an eventual 4x
performance drop relative to the optimal case. The use of the texture cache as a
means of improving memory access has also been tested; overall computation time
for system {\em S} is nearly doubled (actually 1.85x) without the cache,
demonstrating its importance in compensating for the effects of high memory
latency.

\begin{table}
\centering
\caption{Increasing time per atom-step when reordering is omitted.
\label{tab8}}
\vspace{5pt}
\begin{tabular}{rc}
\hline
$N_{step}$ &   $t$    \\
\hline\\[-10pt]
   1000    &   0.088  \\
   2000    &   0.105  \\
   5000    &   0.127  \\
  10000    &   0.151  \\
  20000    &   0.183  \\
  40000    &   0.214  \\
  60000    &   0.232  \\
  80000    &   0.242  \\
 100000    &   0.249  \\
\hline
\end{tabular}
\end{table}

\section{Conclusion}

The present work has been based on a rather modest GPU designed for laptop
computers, whose performance lags behind current high-end devices and, even more
so, behind products scheduled (at the time of writing) to appear in the near
future. Nevertheless, the processing speed is found to be considerably higher
than a typical CPU, a goal achieved without encountering any major algorithmic
or programming obstacles. An especially important feature of the present MD
approach for short-range interactions is that, unlike previous work, it is
completely scalable, enabling it to benefit fully from the inherent parallelism
of the hardware.

It is reasonable to expect that further improvements in GPU design and
performance, particularly the enhanced parallel-processing capability, will
provide a major advance in affordable MD simulation at the single-GPU level for
models of both simple and more complex systems, as well as enabling its
utilization as a convenient building block for massively parallel supercomputers
aimed at extending the realm of feasible simulation.

\bibliography{gpumd}

\bibliographystyle{unsrt}

\end{document}